\documentclass[manuscript]{acmart}

\AtBeginDocument{%
  \providecommand\BibTeX{{%
    \normalfont B\kern-0.5em{\scshape i\kern-0.25em b}\kern-0.8em\TeX}}}

\setcopyright{rightsretained} 
\acmJournal{TMIS}
\acmYear{2021} \acmVolume{1} \acmNumber{1} \acmArticle{1} \acmMonth{1} \acmPrice{}\acmDOI{10.1145/3468877}

\usepackage{csquotes}
\usepackage{pifont}
\newcommand{\cmark}{\ding{51}}%
\begin{document}

\title[Privacy and Confidentiality in Process Mining]{Privacy and Confidentiality in Process Mining - Threats and Research Challenges}

\author{Gamal Elkoumy}
\authornote{Authors in alphabetic order with equal contribution to this research.}
\email{gamal.elkoumy@ut.ee}
\affiliation{%
  \institution{University of Tartu}
  \city{Tartu}
  \country{Estonia}
}

\author{Stephan A. Fahrenkrog-Petersen}
\authornotemark[1]
\affiliation{%
  \institution{Humboldt-Universität zu Berlin}
  \city{Berlin}
  \country{Germany}
}
\email{stephan.fahrenkrog-petersen@hu-berlin.de}

\author{Mohammadreza Fani Sani}
\authornotemark[1]
\affiliation{%
  \institution{RWTH-Aachen University}
  \city{Aachen}
  \country{Germany}
}
\email{fanisani@pads.rwth-aachen.de}

\author{Agnes Koschmider}
\authornotemark[1]
\affiliation{%
  \institution{Kiel University}
  \city{Kiel}
  \country{Germany}
}
\email{ak@informatik.uni-kiel.de}

\author{Felix Mannhardt}
\authornotemark[1]
\affiliation{%
  \institution{Eindhoven University of Technology}
  \city{Eindhoven}
  \country{The Netherlands}
}
\email{f.mannhardt@tue.nl}

\author{Saskia Nuñez von Voigt}
\authornotemark[1]
\affiliation{%
  \institution{Technische Universit\"at Berlin}
  \city{Berlin}
  \country{Germany}
}
\email{saskia.nunezvonvoigt@tu-berlin.de}

\author{Majid Rafiei}
\authornotemark[1]
\affiliation{%
  \institution{RWTH-Aachen University}
  \city{Aachen}
  \country{Germany}
}
\email{majid.rafiei@pads.rwth-aachen.de}

\author{Leopold von Waldthausen}
\authornotemark[1]
\affiliation{%
  \institution{Magdalen College, University of Oxford}
  \city{Oxford}
  \country{United Kingdom}
}
\email{leopold.vonwaldthausen@magd.ox.ac.uk}

\renewcommand{\shortauthors}{Elkoumy, et al.}

\begin{abstract}
Privacy and confidentiality are very important prerequisites for applying process mining in order to comply with regulations and keep company secrets. This paper provides a foundation for future research on privacy-preserving and confidential process mining techniques. Main threats are identified and related to an motivation application scenario in a hospital context as well as to the current body of work on privacy and confidentiality in process mining. A newly developed conceptual model structures the discussion that existing techniques leave room for improvement. This results in a number of important research challenges that should be addressed by future process mining research.
\end{abstract}



\maketitle

\section{Introduction}
Process mining~\cite{DBLP:books/sp/Aalst16} has been successfully applied in analysing and improving processes based on event logs in all kinds of environments. However, the impact of privacy and confidentiality aspects on process mining has received less attention. Both topics are closely related to the responsible application of data science, a topic that has gotten more attention in recent years as data-driven methods start to permeate our society. While privacy concerns informal self-determination, which means the ability to decide what information about a person goes where~\cite{W3ORG}, confidentiality refers to protecting information from an unauthorized disclosure.

A recent study on privacy re-identification risks in event logs showed that there are serious privacy leakages in the vast majority of the event logs used widely in the community~\cite{nunez2020quantifying}. This highlights the need to develop techniques for event logs with an emphasis on privacy and confidentiality preservation. This need was also noted from industry in a recent panel discussion~\cite{DBLP:journals/emisaij/MannhardtKBLTW20}. Although some initial fragmented overview of the research on privacy and confidentiality exist~\cite{mannhardt2018privacy,DBLP:conf/caise/Fahrenkrog-Petersen19}, no comprehensive structuring of the field exists. To address this issue, it is essential to understand what kind of privacy and confidentiality requirements have to be addressed by privacy and confidentiality preserving process mining techniques. 
In this light, our paper provides the following main contributions:
\begin{itemize}
    \item We identify a set of threats for privacy and confidentiality and illustrate them according to an example application scenario (Section~\ref{sec-threats}).
    \item We define a conceptual model for requirements and threats and provide a structured discussion of the current literature (Section~\ref{sec-PPPM}). 
    \item We use the literature review to identify a number of research challenges for privacy and confidentiality in process mining (Section~\ref{sec-discussion}). 
\end{itemize}

\noindent Upfront, we provide a definition of privacy and confidentiality for the purpose of this paper and introduce process mining based on a motivational example scenario.

\subsection{Privacy and Confidentiality}
\label{intro-privacy}

Privacy and confidentiality have a lot in common that may lead to confusion; however, each of them has a specific meaning.

\paragraph{Privacy.} In our current data-driven society, privacy has received much attention through frequent data breaches as well as through regulations such as Europe's General Data Protection Regulation (GDPR) \cite{GDPR}. Generally, privacy is seen as the right of individuals to control how their personal data is collected, used, and/or disclosed to other individuals, organizations or governments~\cite{westin1968privacy}.
GDPR defines personal data as: \enquote{Personal data means any information relating to an identified or identifiable natural person (\enquote{data subject}); an identifiable natural person is one who can be identified, directly or indirectly [..]}\cite{GDPR}. Besides GDPR, privacy is subject to other international laws such as the UN Declaration of Human Rights~\cite{UN}, and Asia-Pacific Economic Cooperation~\cite{APEC}. We follow the definitions of GDPR in this paper.
  
\paragraph{Confidentiality.} Whereas there is some overlap and the concepts are often used interchangeably, the focus of confidentiality is making sure that only authorized individuals have access to the protected data and information. For example, in \cite{harman2012electronic}, confidentiality is defined as an agreement about maintenance and who has access to classified/sensitive data. Thus, confidentiality is concerned with data access, while privacy is focused on individuals and their rights. When the data are \emph{personal data}, the confidentiality challenges coincide with the privacy challenges. Here, we distinguish privacy from confidentiality based on the listed differences. When the main concern is an individuals’ rights, e.g., process workers, customers, or patients, then it is considered a privacy issue. Otherwise, if the concern is more relevant to general data protection, it is assumed as a confidentiality issue. 

\subsection{Process Mining: Preliminaries and Motivational Scenario}
\label{scenarioA}

The foundation of process mining is to use \textit{events} to provide insights about the real execution of business processes. Each event is assumed to be related to certain activities of the process. We introduce process mining and the typical components of a process mining system based on an example scenario in a hospital setting. Health care has seen many process mining applications~\cite{DBLP:journals/tmis/PartingtonWSOK15,Rojas.2016} which makes this a representative example. The goal of analysing processes with process mining in our scenario is to prevent rework, decrease waiting time for patients, and improve documentation by discovering cases of non-compliance. The hospital is also interested in benchmarking, i.e., investigating how their processes and their performance differ from other hospitals.

Concretely, the hospital wants to apply process mining to discover the trajectory of different patients from the moment they are admitted until their discharge from the hospital. Each visit of a patient to the hospital forms a process instance or \emph{case}, and the individual \emph{events} of each case are sourced from the Hospital Information System (HIS). The HIS records information on logistical and treatment activities conducted for specific patients and who of the hospital staff performed them. In addition, a part of the process is performed via e-mails, e.g., referrals to other care institutions and the request of previous medical documentation. Therefore, certain events are collected from the e-mail server of the hospital. E-mails are associated with certain process activities using text mining and the metadata (sender, recipient) is used to identify the staff responsible. Finally, to benchmark the hospital wants to share some of this data over organizational boundaries and perform inter-organizational process mining~\cite{SCHULZ2004109,ZENG20131280}. 

Based on this scenario, we describe the elements of such a typical process mining application. We organize the elements into three layers: data, application, and presentation layer.

\begin{table}[tb]
\centering
\caption{A fragment of a simplified event log obtained from a hospital: each row corresponds to an event.}
\label{tab: eventLog}
\begin{tabular}{|l|l|l|l|l|l|l|} 
\hline
\textbf{Patient} & \textbf{Activity} & \textbf{Timestamp} & \textbf{Resource} & \textbf{Role}  & \textbf{Age} & \textbf{Disease} \\ 
\hline
\textbf{1}       & Register          & 07.01.2020-08:30   & ResA              & Administrative & 22           & Flu              \\
\textbf{1}       & Visit             & 07.01.2020-08:45   & ResB              & Doctor         & 22           & Flu              \\
2                & Register          & 07.01.2020-08:46   & ResA              & Administrative & 30           & Infection        \\
3                & Register          & 07.01.2020-08:50   & ResA              & Administrative & 32           & Infection        \\
\textbf{1}       & Blood Test        & 07.01.2020-08:57   & ResE              & Administrative & 22           & Flu              \\
\textbf{1}       & Discharge         & 07.01.2020-08:58   & ResC              & Administrative & 22           & Flu              \\
2                & Hospitalize       & 07.01.2020-09:01   & ResD              & Administrative & 30           & Infection        \\
3                & Hospitalize       & 07.01.2020-10:00   & ResD              & Administrative & 32           & Infection        \\
2                & Blood Test        & 07.01.2020-10:02   & ResE              & Nurse          & 30           & Infection        \\
3                & Blood Test        & 07.01.2020-10:15   & ResE              & Nurse          & 32           & Infection        \\
2                & Blood Test        & 07.02.2020-08:00   & ResE              & Nurse          & 30           & Infection        \\
2                & Visit             & 07.02.2020-09:30   & ResF              & Doctor         & 30           & Infection        \\
3                & Visit             & 07.02.2020-13:55   & ResF              & Doctor         & 32           & Infection        \\
2                & Discharge         & 07.02.2020-14:00   & ResG              & Administrative & 30           & Infection        \\
3                & Discharge         & 07.02.2020-14:15   & ResG              & Administrative & 32           & Infection        \\
\hline
\end{tabular}
\end{table}

\paragraph{Data Layer.} Process mining starts from event data, a collection of events or event log representing the execution of several instances of a business process. Table~\ref{tab: eventLog} shows an example of such an event log extracted from the HIS and e-mail system of the hospital. Here, each row represents an event that indicates when an activity was performed (\textit{Timestamp}) and by whom it was performed (\textit{Resource}). Furthermore, each event is associated with a running process instance or case (\textit{Patient}), which in our case is a unique identifier of the patient visit. By grouping events based on their case and ordering them according to their timestamp, we obtain sequences of events: one \emph{trace} for each process instance. For example, in Table~\ref{tab: eventLog}, the trace for case $1$ is the sequence of activities: \textit{Register}, \textit{Visit}, \textit{Blood Test}, and \textit{Discharge}. In addition event logs often include additional domain specific attributes, which are not strictly required for the application of basic process mining techniques, but may provide additional context. In our hospital scenario, these attributes are \textit{Age} and \textit{Disease}.

\begin{figure}[tb]
    \centering
    \includegraphics[width=\columnwidth]{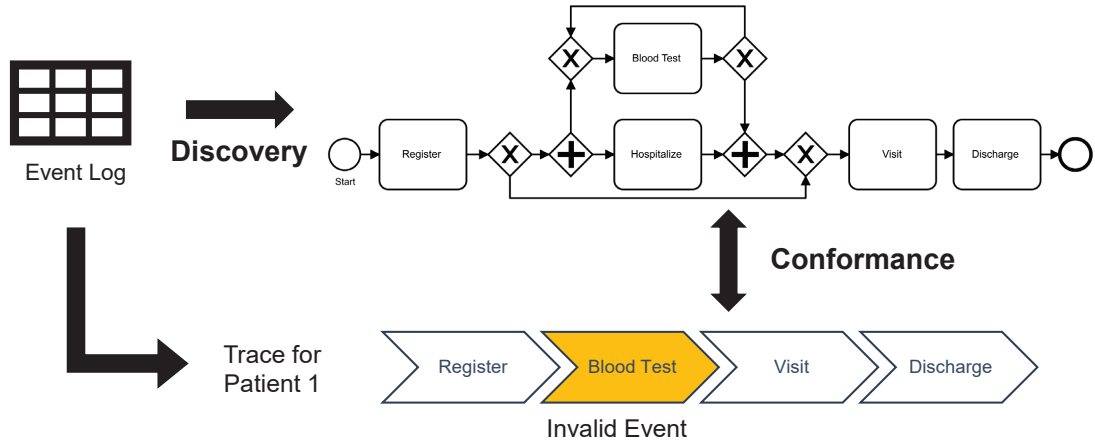}
    \caption{Process discovery and conformance checking shown based on the event log from Table \ref{tab: eventLog}. A process model in BPMN was discovered and used for conformance checking towards the first trace of the log. One of the events of the trace (\emph{Blood Test}) should not have occurred according to the model.}
    \Description{The figure shows how process discovery and conformance checking are done based on an example event log. A BPMN process model consisting of the activities Register, Blood Test, Hospitalize, Visit, and Discharge is discovered.}
    \label{fig:processModel}
\end{figure}

\paragraph{Application Layer} Algorithms that process event logs and compute representations are an important part of process mining. Two of the most important applications of process mining are \textit{process discovery}~\cite{DBLP:journals/tkde/AugustoCDRMMMS19} and \textit{conformance checking}~\cite{DBLP:books/sp/CarmonaDSW18}. Process discovery receives an event log and returns a process model that describes the process behavior in an abstract model defining the possible sequences of activities. An example of such model that could be discovered from the log in Table~\ref{tab: eventLog} would be the BPMN\footnote{Business Process Model and Notation (BPMN) standard: \url{https://www.omg.org/spec/BPMN/}.} model shown in Figure~\ref{fig:processModel}. Here the process discovery algorithm inferred from the event log that each trace starts with activity \emph{Register}. Then, activity \emph{Hospitalize} must be performed in parallel with one or multiple \emph{Blood Tests}. Finally, each case of our simplified process is concluded by a \emph{Visit} in sequence with \emph{Discharge}. Conformance checking aims to quantify deviations between the process model and real execution data as observed through the event log. The output of a conformance checking algorithm is usually information about how individual traces in the given event log deviate from that reference model. In Figure~\ref{fig:processModel}, a conformance checking technique has identified that the activity \emph{Blood Test}, which occurred according to an event in the trace for patient 1, should not have been performed since the patient was not hospitalized. Many other tasks are possible such as the mining of resource profiles of the employees~\cite{DBLP:journals/tmis/PikaLWFHA17}, the mining of decision rules~\cite{DBLP:books/sp/Aalst16} based on the characteristics of cases captured in additional attributes, or the automated prediction of the next step in a running case~\cite{10.1145/3301300} with the goal of acting on cases leading to a bad process outcome or performance.

\begin{figure}[tb]
    \centering
    \includegraphics[width=\columnwidth]{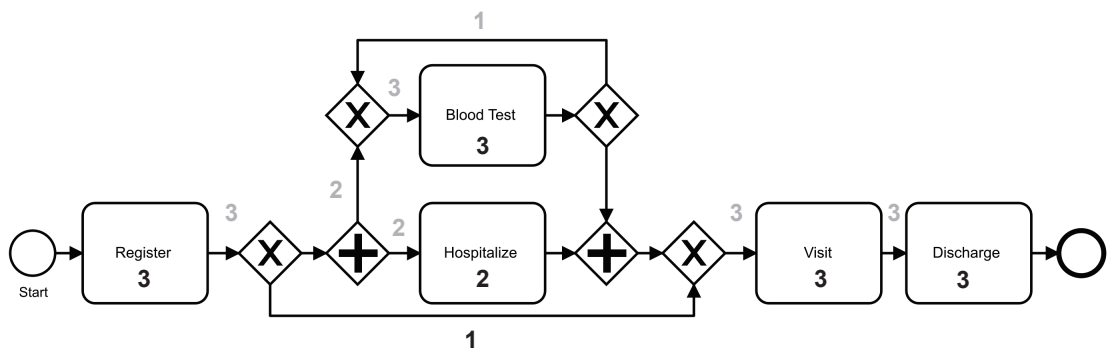}
    \caption{A frequency annotated process model based on the combined results of process discovery and conformance checking. When only considering the presentation layer of the process mining application, the analyst may only see such a visualization of the underlying event log. Occurrence frequencies of activities (in black) and of the transitions between activities (in grey) are added as annotation to the model.}
    \Description{The figures shows the same process model as in the previous figure. However, here the frequencies based on the event log are added as annotation on the activities and edges.}
    \label{fig:processModelPresentation}
\end{figure}

\paragraph{Presentation Layer.} Generated artifacts such as the discovered process models or the deviations detected by conformance checking, or other analytical outputs need to be presented to the process analyst or business user. Often these outputs are aggregated representations, e.g., a process model with projected frequencies as shown in Figure~\ref{fig:processModelPresentation}. In many cases, these representations also include additional information, e.g., the average waiting or processing times of activities. Regarding privacy and confidentiality, it is important to note that these artefacts do not directly reveal the exact underlying event logs and only provide an aggregated view on the process reality. In many cases, process analysts or domain users could use these refined results to obtain insights on the process without getting access to the underlying event data or algorithmic results.  
However, also in the presentation layer, results can be specific to single events or cases. They may be infrequent paths in a frequency annotated process model, e.g., the path from \emph{Register} to \emph{Visit} annotated with frequency 1 in Figure~\ref{fig:processModelPresentation}. Many process mining tools also provide the option to drill down to, e.g., a single process prediction, a compliance violation, or simply give direct access to the source events.

We briefly introduced the major elements of process mining relevant to the discussion of privacy and confidentiality threats along with three layers. In the following sections, we will show how these elements are related to the design of privacy-preserving and confidential process mining techniques. 

\section{Threats for Privacy and Confidentiality in Process Mining}
\label{sec-threats}
This section presents threats to confidentiality and privacy in process mining. We collected the threats based on, first, developing a list of concrete attacks among the authors and, then, cross-referencing this list with generic attacks from the literature.

In this paper, we do not focus on attacks with malicious adversaries that control the information flow. We assume an honest-but-curious attacker who follows the protocol and has legitimate access to the data. An attacker might be the process analysts or business user who obtains sensitive information from supposedly anonymized event data provided to her. Inside or outside attacks to bypass access control are outside our scope. 

Overall, the identified threats can be categorized into four categories: re-identification, reconstruction, membership disclosure, cryptanalysis; each of which is described in more detail based on the literature and is instantiated in the context of our hospital scenario.

\subsection{Re-identification Threats (T1)}
Re-identification or de-anonymization threats are those where the identity of an information (data) subject is at risk to be disclosed by singling-out individuals from the supposedly anonymized event logs~\cite{dwork17}. This threat of re-identification is currently most dominant at the data layer. Here an information subject may be directly linked to the process case, e.g., the patient in Table~\ref{tab: eventLog}, or linked to a certain activity, e.g., a resource (employee) of the hospital.
There are several possible attack methods described in the literature that constitute re-identification threats. 

In a \textit{linkage attack}, an attacker uses background or context knowledge on the process or on the individuals and combines it with the released artefact. A pseudonimized event log in the data layer of the process mining application may be such an artefact to which an analyst, or the general public in case of research data, has access. At first glance, Table~\ref{tab: eventLog} does not allow to re-identify patients or employees, all direct identifiers have been replaced. However, based on equal unique attributes, events can be linked to a case and thus to an identity. 
For instance, assume an attacker knows that Alice is 32 years old and has visited the hospital in a certain timeframe. By linking this very basic information with Table~\ref{tab: eventLog} only patient 3 is 32 years old, an attacker can infer the corresponding events of this patient and the sensitive disease. Such linkage attack can also be based on unique combinations of activities that are performed in a sequence for a certain process case. For example, the process case for the patient with identifier 2 is the only one containing two occurrences of the activity \emph{Blood Test}. Thus, the uniqueness of cases in an event log can indicate the risk of re-identification. It was shown in~\cite{nunez2020quantifying} that there is serious potential for privacy leaks in published event log data, as the vast majority of public research event logs contain many unique cases.

When several organizations independently release generalized event logs about overlapping populations, re-identification is possible by an \textit{intersection attack}. This may happen in the inter-organizational process mining setting as illustrated with the benchmarking use case in our hospital example or, e.g., when several government agencies release event log data, which is likely to be containing information about the same population of citizens. If we assume that an adversary knows that a target is contained in several event logs, the identity may be disclosed by taking the intersection. Assume two hospitals independently publish event logs of patient trajectories in which age is generalized to prevent a linkage attack, i.e., Table~\ref{tab: eventLog} would only contain age groups 0--20, 21--40, and so on. Patient Alice would not be easily re-identifiable anymore. However, let us assume that an adversary knows that she was transferred from hospital A to hospital B. Even though the age group is generalised, there may only be one process case in that age group in each of the hospital event logs such that it is consistent with the transfer scenario. So, the intersection set is a single record and we have re-identified Alice. Of course, such an attack may also be performed on other event log attributes and not only be based on the time relation between process cases. 



\subsection{Reconstruction Threats (T2)}
The threat of reconstruction is the risk of recreating the (partial) original 
event log from a released process model or aggregated 
statistics~\cite{DinurN03}. Reconstruction is a threat to privacy when attributes from individuals are reconstructed, and a threat to confidentiality when reconstructing non-personal data. This threat is closely linked to the presentation layer of the process mining application. The individuals in the event log are seemingly protected by only releasing aggregate statistics. 

If an individual cannot be linked directly, attributes can be reconstructed, for example, from aggregated statistics by a \textit{difference attack}. 
In this attack, an adversary isolates a single value by combining multiple aggregated statistics about a dataset. Assume the process analyst can only pose queries to obtain aggregate statistics, e.g., by obtaining process model visualizations such as the one in Figure~\ref{fig:processModelPresentation}. The analyst could obtain the frequency visualization grouped per disease and, therefore, know the number of patients (unique cases) for each disease. In a second query, the analyst could exclude 32-year-old patients in the query and obtain the same statistics. From the difference, an adversary can infer that Alice, the only 32-year-old patient, has an infection. 

Adversaries may also attempt to reconstruct training or source data from a published model, which is called \textit{model-inversion}~\cite{FredriksonJR15}. The training data is estimated by observing the input and output of a model. This attack only creates a probabilistic version of the training data. The models used for predictive and prescriptive process monitoring~\cite{fahrenkrog2019fire,maggi2014predictive} use machine learning models that could be directly vulnerable to this attack. In our application scenario, however, even a probabilistic version of the original event log can reveal private information such as the diagnosis of a specific patient or the responsible resources treating a patient. While model-inversion attacks have not yet been well studies for process mining, for instance, from the process model in Figure \ref{fig:processModelPresentation}, we can infer that the underlying training event log consists of a single trace in which the patient left without being hospitalized and also that there is a single patient that had two blood tests taken.  

\subsection{Membership Disclosure Threats (T3)}
Membership disclosure threats entail uncovering the knowledge of whether a specific individual was included in the source data for a particular model or analysis. So, differently to a re-identification attack only the fact that an individual was part of the dataset is disclosed. 

In a \textit{membership inference attack}, an adversary aims to determine whether an individual was included in the source data. An adversary who only has access to the released model can train shadow models to predict membership \cite{ShokriSSS17}. These models capture the misclassification difference between samples that are likely to be in or outside the training data. By checking if a process model allows the behavior of a certain trace, an adversary can try to predict, if the trace was included in the data underlying the process discovery. In our hospital example, depending on the scope of a process mining analysis, such knowledge could reveal that a specific individual visited the hospital for a specific treatment. For instance, assume an attacker knows that a target patient Bob has made a blood test twice and has access to the process model (cf. Figure~\ref{fig:processModelPresentation}). It is very likely that Bob is included since he made a blood test twice. However, in this case, an adversary cannot identify that Bob is patient 2, i.e., the exact process case identifier remains protected. Still, this knowledge may leak sensitive information, e.g., if the dataset was extracted for a set of patients with a specific disease membership disclosure would also disclose the disease information.

\subsection{Cryptanalysis Threats (T4)}
Often, data is pseudonymized, as in our example, or event encrypted to provide confidentiality. However, pseudonymized or even fully encrypted event logs are vulnerable to attacks based on the analysis of the frequency.

A \textit{frequency analysis} takes advantage of the characteristics of the encrypted data. Such analysis could rely on background knowledge of the process, e.g. the frequency of certain activities within one trace and their position in the trace. For example, when considering our hospital process example, certain diagnostic steps, such as blood test, might appear more than once in a trace, while the registration and release of a patient usually happen at the beginning and end of each trace. In other words, even if the activity in Table \ref{tab: eventLog} had been encrypted an attacker might decode the start and end activity. In this case, an attacker has both the plaintext and its encrypted version, which can be used to reveal the total cipher. This is also possible for the resource. Here, the adversary's background knowledge may be that, for example, only \textit{Natsa} was working on the registration on January 07, 2020. In this case, an adversary can link the activities \textit{Register} with the pseudonymized resource identifier \textit{ResA}. It is not difficult to gain this knowledge, especially in public places like a hospital. 

We identified four main threats and sketched the related attack methods in a process mining scenario. In the remainder of this paper, we review and discuss possible solutions and identify the open challenges.

\section{Privacy-preserving Process Mining}
\label{sec-PPPM}
In this section, we discuss requirements that address the threats discussed in Section~\ref{sec-threats}. The requirements are taken from a systematic synthesis of the current privacy and confidentiality landscape conducted by Gharib et al. \cite{Gharib/CoPri2020}, who themselves based their work on a previous literature review \cite{Gharib/ER2017}. The mentioned requirements are legislature agnostic but nonetheless present the opportunity to incorporate demands and elements of multiple common protection models such as the European (GDPR), Australian (Privacy Act 1988), Canadian (PIPEDA), and US legislation. We will particularly focus on GDPR as an example to explain the origin of the requirements.

\subsection{Conceptual Model and Requirements}

\begin{figure}[tbp]
    \centering
    \includegraphics[width=0.7\columnwidth]{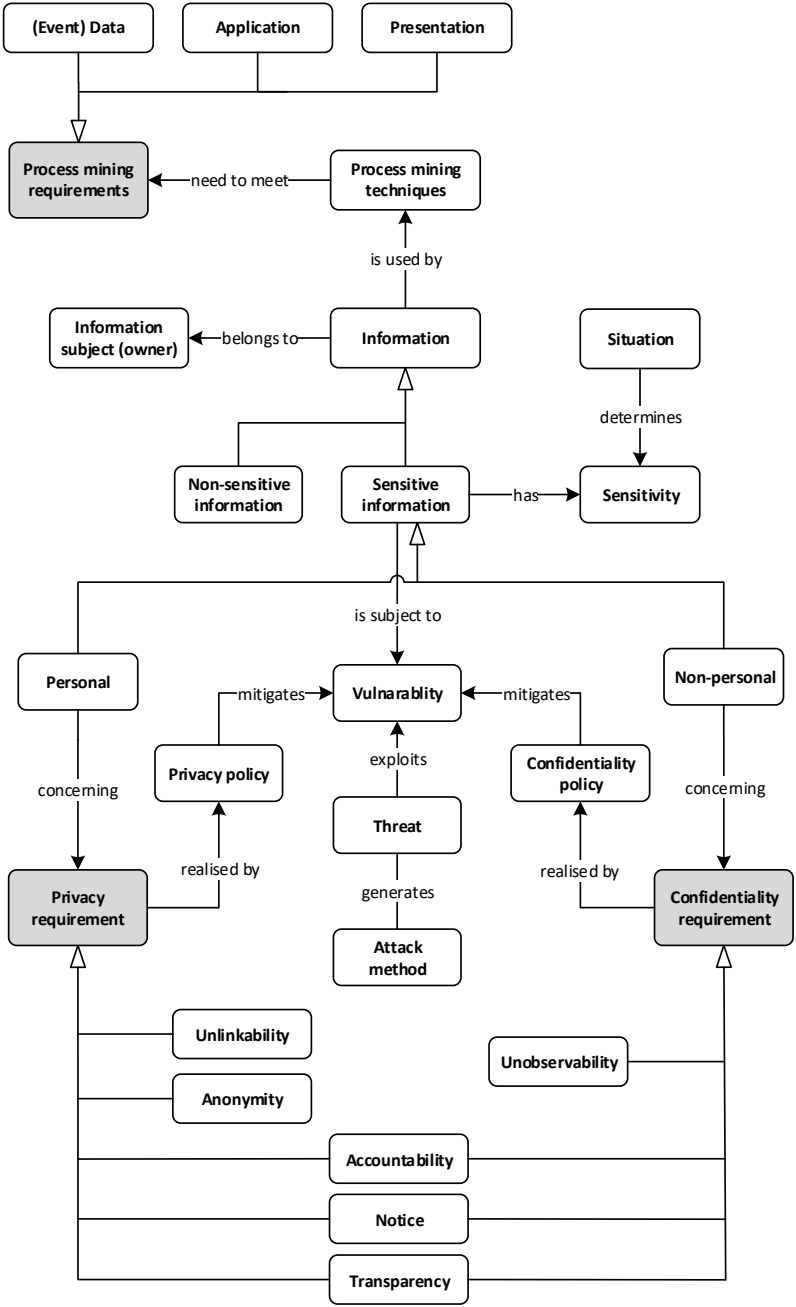}
    \caption{A UML class diagram linking threats and requirements to the remaining concepts used.}
     \Description{The figure shows a UML class diagram that relates the concepts of privacy and confidentiality to the respective requirements and the type of information that is used: personal or non-personal information.}
    \label{fig:PCModelExt}
\end{figure}

Figure~\ref{fig:PCModelExt} shows a UML class diagram of how the concepts of privacy and confidentiality relate, both to the discussed threats and to the requirements taken from \cite{Gharib/CoPri2020}. In Section~\ref{intro-privacy}, we note that confidentiality has direct overlap with privacy. Therefore, confidentiality could, as in Gharib et al. \cite{Gharib/CoPri2020}, solely be seen as a requirement of privacy. However, following our definition in Section~\ref{intro-privacy}, we raise confidentiality beyond just its overlap with privacy and make it a separate concept, therefore slightly adjusting the model in Gharib et al. \cite{Gharib/CoPri2020}. The purpose of our privacy and confidentiality requirements is to address potential threats and mitigate potential vulnerabilities. In the following, we explain the requirements.

\begin{itemize}
   \item \textbf{R1 - Anonymity} deals with personal information, and it ensures that personal data can only be used without disclosing identities of information subjects~\cite{teminology_2006_dritsas,PfitzmannK00}. This is a privacy requirement since it concerns personal information. According to Recital 26 of GDPR, the principles of data protection should not apply to anonymous information where the information subject is not identifiable. Therefore, anonymity is a big step towards privacy regulations compliance.  
   \item \textbf{R2 - Unlinkability} describes that it should not be possible to link personal information to their corresponding owners~\cite{PfitzmannK00}. This requirement complements R1 in the sense that preventing identity disclosure cannot be guaranteed by only making identifiers unreadable. Furthermore, all identifiers for linkage also need to be removed. Unlinkability includes that data cannot be re-identified by linkage attacks.    
   \item \textbf{R3 - Unobservability} means that it should not be possible to observe the identities of information subjects that perform any action~\cite{PfitzmannK00}. It should be noted that unlike anonymity and unlinkability, which keep the identity of the actor hidden, the goal here is to ensure that the actions themselves are hidden. Thus, we categorize this requirement as a confidentiality requirement. This requirement mostly concerns continuous monitoring and processing of the data generated by running systems. It is intended to protect personal data against unauthorized processing, as described in Article 5 of GDPR.     
   \item \textbf{R4 - Notice} means that information subjects should be notified when their information is gathered~\cite{teminology_2006_dritsas}. A notice should include the detailed information which is to be gathered, disclosure risks and data quality concerns. The notice requirement can be categorized as both privacy and confidentiality requirement depending on the information owner, which could be both an individual or a company. This requirement relates to the concept of \textit{consent} in GDPR. The data processing often needs to be based on consent. As mentioned in Article 7 of GDPR, the data controller/processor should demonstrate that the information subject has consented to processing of his/her data. Note that the consent needs to be kept updated based on the purpose of data processing.   
   \item \textbf{R5 - Transparency} means information owners should be able to know who uses their information, how, and for what purposes \cite{teminology_2006_dritsas}. This requirement can also be categorized as both privacy and confidentiality requirement depending on the nature of the information owner. The principle of transparency is described in Recital 58 of GDPR.
   \item \textbf{R6 - Accountability} describes that information subjects should be able to hold information users accountable for their activities and the consequences of misusing their data \cite{teminology_2006_dritsas}. This requirement concerns both personal and non-personal information. It is mentioned in Article 5 GDPR as one of the principles of processing personal data, where it is described that the personal data should be protected against unlawful processing, accidental loss, destruction, or damage.
\end{itemize}

When information is exploited by process mining techniques, protecting privacy and confidentiality should not compromise the requirements of process mining techniques. There are three types of additional requirements:

\begin{itemize}
    \item \textbf{R7 - Data} requirements are that process mining techniques should support different data storage formats, e.g., centralized in a single organization and distributed among different parties. On top of that, processing different types of data should not lead to any privacy leakage.
    
    \item \textbf{R8 - Application} requirements mean that the algorithms which are applied should be computationally acceptable, and fulfilling privacy and confidentiality requirements should not impose an unreasonable load on the time or resource consumption of the algorithms.
    
    \item \textbf{R9 - Presentation} requirements are that the reported results should be interpretable by users. This includes fulfilling privacy and confidentiality requirements without leading to a utility loss of the anonymized data and having the ability to repeat different types of queries without privacy disclosure.
\end{itemize}

\subsection{Existing Protection Models}

Here, we introduce the main strategies of existing protection models~\cite{wagner2018technical} that are designed to provide technical solutions for the requirements. There are three main categories:

\begin{itemize}
    \item \textbf{(M1) Group-based models} based on data similarity like $k$-anonymity \cite{sweeney2002k} and its extensions, e.g., $l$-diversity \cite{machanavajjhala2007diversity} and $t$-closeness \cite{li2007t}. 
    These types of models are aimed to provide \textit{R1 - Anonymity} and \textit{R2 - Unlinkability} requirements.
    Here, cases are grouped such that each case shares the quasi-identifier values with a group of cases. However, this protection-based model makes assumptions about the background knowledge of an adversary.  
    \item \textbf{(M2) Indistinguishability-based models} introduce noise to provide uncertainty of whether an individual's data is included in the dataset. The best-known models comply with differential privacy~\cite{dwork2008differential} as a provable guarantee to ensures that removal or addition of a single case does not significantly impact the result of analysis regardless of an adversary's knowledge. This type of models is also focused on \textit{R1 - Anonymity} and \textit{R2 - Unlinkability} requirements.
    \item \textbf{(M3) Confidentiality frameworks}~\cite{rafieiWA19_short} are often based on encryption methods and access control requirements, where different encryption techniques are applied to data based on their sensitivity and the authorization of users who need to access the data. Confidentiality frameworks could support \textit{R1 - Anonymity}, 
    \textit{R3 - Unobservability}, and \textit{R5 - Transparency} requirements. 
\end{itemize}

Additional orthogonal models, such as those based on the idea of information gain~\cite{kenthapadi2005simulatable} or time~\cite{wright2002analysis}, are important for confidentiality and privacy alike.

\section{Research Challenges}
\label{sec-discussion}

Based on the identified threats, requirements, and protection models, we review the literature and synthesise a set of research challenges. 

\subsection{Current work}
Previous studies on privacy-preserving and confidential process mining addressed some of the requirements and attempts to mitigate the threats. 

Rafiei et al.~\cite{rafiei2020tlkc} propose a privacy model called TLKC for publishing event logs, which provides group-based anonymization (M1) based on k-anonymity, and quantifies the risk based on the attacker's background knowledge. Their model fulfills R1 - Anonymity and R2 - Unlinkability). The k-anonymity model is secure against the reconstruction threat (T2). However, k-anonymity partially mitigates the re-identification threats (T1) and the membership disclosure threat (T3)~\cite{cohen2020towards}. The adoption of k-anonymity makes the model interpretable (R9 - Presentation). 

Also, Fahrenkrog-Petersen et al.~\cite{fahrenkrog2019pretsa,bauer2019elpaas} use the k-anonymity privacy model (M1) for publishing event logs but add the use of t-closeness to protect attribute values.
Similar to TLKC, R1 - Anonymity, R2 - Unlinkability requirements are fulfilled. Again, the privacy model is interpretable (R9 - Presentation). The authors suppress events or cases that would result in privacy leakage. However, data suppression is correlated with utility loss and no explicit utility measure is considered.

The differential-privacy model (M2) is adopted, e.g., by Mannhardt et al.~\cite{mannhardt2019privacy} for controlling disclosure of two types of queries: the frequencies of directly-follows relations and the trace variant frequencies. Re-identification threats T1, reconstruction threats T2, and membership disclosure threats T3 are mitigated and requirements R1 and R2 are fulfilled. Neither transparency nor accountability is considered, and no guarantees are given for the utility (R9 - Presentation) of the obtained DFG.


Fahrenkrog-Peterson et al.~\cite{fahrenkrog2020pripel} use a differential privacy mechanism as well for event log anonymization in a framework called PRIPEL. They ensure privacy guarantees on the basis of individual cases. Also, they consider anonymizing event log attributes with different $\epsilon$ values, adapted with respect to the sensitivity of their values. Furthermore, they propose timestamp shifts to anonymize the timestamp attribute. The use of differential privacy fulfills the requirements R1 - Anonymity and R2 - Unlinkability to mitigate threats T1, T2, and T3. However, PRIPEL does not optimize the disclosure for a certain level of utility (R9 - Presentation).

Cryptographic privacy models (M3) have been considered for both centralized and distributed event logs settings. Rafiei et al.~\cite{rafieiWA19_short,rafiei2018ensuring} introduced an encryption framework for ensuring confidentiality in process mining to secure against the threat T4. The framework is divided into three processing environments and provides user access control thereby addressing R6 - Accountability. The framework gives a data analyst access to the internal partially secure event logs, which makes the framework vulnerable against T1, T2, and T3. Only centralized event logs are considered failing to provide support for distributed event logs (R7 - Data). A cross-organizational setting is considered, e.g., by Tillem et al.~\cite{tillem2017mining} who propose a secure processing protocol to execute the Alpha algorithm over distributed event logs (R7 - Data). However, their protocol does not mitigate attacks on confidentiality described in T4.

Elkoumy et al.~\cite{elkoumy2020secure,elkoumy2020shareprom} adopt secure multi-party computation protocols to jointly calculate directly-follows relations securely between several organizations without the need to share private data in order to fulfill R7 - Data. Their framework is secure against the attacks on confidentiality T4. They added a differential-privacy layer~\cite{elkoumy2020shareprom} in order to mitigate the threat on privacy T1 to fulfill the requirement R1 - Anonymity. However, the effect on the utility loss has not been studied (R9 - Presentation).

Other studies on privacy-preserving process mining do not fulfill any of the above requirements as they do not provide a concrete mechanism of disclosure control. Rafiei et al. \cite{rafieippdp_google,rafieippdpTool_short} provide privacy metadata by extending the XES standard. Pika et al. \cite{pika2020privacy} studied the impact of anonymization on process mining in healthcare without providing a concrete mechanism. In this line, Rafiei et al. \cite{rafiei2020towards} provide privacy quantifications for both the disclosure risk and the utility loss and Nu{\~n}ez von Voigt et al. \cite{nunez2020quantifying} quantify the re-identification risk resulted from the disclosure of event logs based on individual uniqueness. Both do not provide a solution. Finally, other works offer only one specific task, for example, Rafiei et al.~\cite{RafieiA19} provide privacy-preserving role mining adopting a substitution method that secures the activities with sensitive frequencies.

We summarize the studies in Table~\ref{tbl:related_work} and observe that most of the previous studies fulfill the requirements: R1, R2, and R3. However, some of the requirements have not been addressed in the literature, e.g., R4, R5, R8, and R9. Furthermore, the literature either mitigates threats on privacy or threats on confidentiality, but it does not mitigate both types of the attacks together.

\begin{table}[hbtp]
	\centering	
	\caption{Summary of privacy-preserving process mining approaches w.r.t. the requirements, the protection models, and the threats (the symbol \cmark means fulfillment, $\ast$ means partial fulfillment, and - means does not fulfill)}
	\begin{tabular}[t] { |p{2.8cm}||c|c|c|c|c|c|c|c|c||c|c|c||c|c|c|c|}
		\hline
		
		Paper     &R1&R2&R3& R4&R5&R6&R7&R8&R9& M1&M2&M3 & T1&T2&T3&T4\\\hline 
		    TLKC~\cite{rafiei2020tlkc} &\cmark & \cmark& - & - & - & - & - & - 
		    &\cmark&\cmark & - & - & $\ast$ &\cmark&$\ast$&- \\\hline
		    
		    Fahrenkrog-Petersen et 
		    al.~\cite{fahrenkrog2019pretsa,bauer2019elpaas} &\cmark & \cmark& 
		    -&-&-&-& -&-&\cmark&\cmark&-&-&$\ast$&\cmark&$\ast$&-\\\hline
		    
		    Mannhardt et al.~\cite{mannhardt2019privacy} &\cmark & \cmark& 
		    -&-&-&-& -&-&-&-&\cmark&-&\cmark&\cmark&\cmark&-\\\hline
		    
		    PRIPEL~\cite{fahrenkrog2020pripel} &\cmark & \cmark& -&-&-&-& 
		    -&-&-&-&\cmark&-&\cmark&\cmark&\cmark&- \\\hline
		    
		    Rafiei et al.~\cite{rafieiWA19_short,rafiei2018ensuring} &- & 
		    -& \cmark&-&-&\cmark& -&-&-&-&-&\cmark&-&-&-&\cmark\\\hline
		    
		    Tillem et al.~\cite{tillem2017mining}&- & -& -&-&-&-&\cmark&-& 
		    -&-&-&\cmark&-&-&-&- \\\hline
		    
		    Elkoumy et al.~\cite{elkoumy2020secure,elkoumy2020shareprom} 
		    &\cmark & -& -&-&-&-& 
		    \cmark&-&-&-&-&\cmark&\cmark&-&-&\cmark\\\hline
	\end{tabular}
	
	\label{tbl:related_work}
\end{table}

\subsection{Research Challenges for Process Mining}
Our analysis of state-of-the-art work addressing privacy and confidentiality gives rise to a number of research challenges.

\begin{enumerate}
\item \textbf{Interpretable Quantification of Privacy Disclosure}
In real life, there is always privacy loss with any information disclosure. Organizations need more reliable and interpretable metrics of privacy disclosure (R9 - Presentation). Interpretable disclosure metrics of the applied privacy models are needed that a user can understand in business terminology \cite{wagner2018technical}. Translated to our hospital example, the managers need risk indicators that they can understand. A probabilistic percentage indicator may help address this challenge as they are easier to understand by end-users.

Such risk metric would need to consider disclosure at several levels and under different assumptions: the entire event log, specific cases or events, random cases or events, multiple or single cases per individual. Specifically, how many cases are recorded for a single individuals is an important factor for M3 models~\cite{DBLP:journals/tmis/KartalLL19}. Concretely, the hospital in our example may have a highly sensitive patient (e.g., a politician) that has higher privacy requirements than the other patients, or a specific treatment (activity) can be riskier than the other treatments, e.g., treatments directly related to rare diseases. Personalized privacy quantification may help to address this challenge. The $\epsilon$-value for differential privacy, often guaranteed by M3 models, may be used as indicator, but is hardly interpretable by the general public. K-anonymity based models such as~\cite{rafiei2020tlkc} are interpretable, but only partly mitigate some threats.

\item \textbf{Balancing Risk and Utility}
The perfect way to protect private data is to avoid sharing it. However, insights are being missed when necessary data is not available. Thus, there is always a trade-off between the disclosure risk and the utility of using the disclosed data \cite{lee2011much,hsu2014differential}, as discussed in R9. Hence, it is a challenge to perform privacy-preserving process mining without losing the utility of the anonymized event log. In our example, hospital managers need to estimate the amount of noise to be added to an event log to achieve a certain level of utility with acceptable risks. This estimation depends highly on the kind of analysis that is to be performed. Moreover, in line with the concepts of \emph{consent} in GDPR, it is important to decide upfront on what risk can be accepted. However, process mining is often used for exploratory analysis in the hope to identify patterns in the event data, which makes this trade-off challenging. 

\item \textbf{Level of Granularity}
Process mining tools enable the user to perform analysis from different perspectives, e.g., activity-centric, to study the directly-follows relations, and resource-centric to study the hand-offs between resources. Moreover, other types of analysis, such as task mining enable the assessment of specific processes and offer their best automation opportunities. For instance, the hospital can apply a privacy model to the activity perspective; however, this does not protect the disclosure of how many resources are being utilized per hour. The privacy of patients would be protected, but events could be used for work surveillance. A clear research gap is that current work focuses on specific perspectives and attack scenarios and, due to the richness of event data, many threats and perspectives, e.g., the work surveillance perspective, have not yet been addressed.

Granularity is also an opportunity. Many process mining tools enable the user to control the level of abstraction and apply filters on the discovered process model in order to fulfill the presentation layer requirement R9. Hence, such dashboards and business process filtering \cite{zaman2019process} could be provided to the process analyst while achieving an acceptable level of privacy. For instance, the hospital may decide to disclose the process map with the most frequent 10\% of edges because that implies lower risk. Research on such systems and scenarios is currently missing.

\item\textbf{Distributed Privacy}
Distributed privacy-preserving process mining aims to get insights from several event logs originating from several business organizations with a disclosure control mechanism. Different parties could be competitors and may not want to provide their full local private data to one another. However, they expect a mutual benefit by analysing global insights with process mining. 
For example, two hospitals may share mutual patients and they need to optimize their inter-organizational processes. Furthermore, an inter-organizational conformance checking or inter-organizational variant analysis may be needed to fulfill new business collaboration needs.
A common solution to this problem is security protocols. However, such techniques have high latency and massive communication overhead and require special deployments among organizations~\cite{elkoumy2020secure}. There is still a gap between the developed techniques and their real-life usability, also evident from the complete lack of related business studies for process mining. 

\item \textbf{Computational Challenges}
The optimal k-anonymization is NP-hard~\cite{meyerson2004complexity}. Hence, with increasing the dimensionality of the attributes of the event log, it becomes more unpractical to achieve privacy-preserving process mining. Similar arguments hold for both t-closeness and l-diversity models. Therefore, achieving an optimal privacy-preserving process mining with sufficient execution time (R8 - Application) is challenging. We need more efficient and scalable computational models for privacy-preserving process mining. Often, privacy and confidentiality is an afterthought since the value-adding business requirements are considered first. This, makes is imperative to provide techniques that are usable in practice.

\item \textbf{Traceability Challenge}
Due to privacy regulations, such as GDPR, it is essential for an organization to provide individual to provide consent (right to consent), to access (right to access), and to remove (right to be forgotten) their personal data. Hence, organizations need to trace the data life-cycle starting from the data collection to the data removal from their databases~\cite{zaman2019process}. There is an inherited challenge to provide traceability, primarily when data is distributed across information systems, specifically with process mining, where analysts look at processes across silos. While the application of process mining to verify the compliance with privacy regulations has been reported~\cite{DBLP:journals/sncs/ZamanH20}, there has not yet been research on the usage of recorded data for the purpose of process mining, which is often a secondary use of already collected data.

\item \textbf{Transparency Challenge.}
According to regulations, the information subject should be notified when their data is used (R4 - Notice), and they should be able to know who is using the data and for what purposes (R5 - Transparency). Techniques need to be developed to support transparency, such as authentication and authorization, and audit-log mechanisms. Audit-log mechanisms record information such as the involved user, date and time, and the action executed by the user. This includes that organizations inform the information subjects when and for what purpose their data was used. Full traceability is a requirement for achieving transparency, but not sufficient by itself. Event logs are often collected from distributed data sources making it challenging to achieve transparency. Research on such mechanisms and systems is missing in the process mining domain.
\end{enumerate}

\section{Conclusion}
\label{sec-conclusion}

According to regulations, organizations are obliged to adhere to privacy compliance and responsible usage of data. Additionally, confidentiality is of great importance in any professional setting. This poses challenges that require extensions of current state-of-the-art process mining techniques, which are often applied to potentially sensitive data. Designing appropriate techniques requires knowing the existing threats and challenges. This paper presents a conceptual model for threats and requirements that must be fulfilled by privacy and confidentiality preserving process mining techniques. Our literature survey highlights a need for future techniques that comprehensively address those threats and requirements.

In the future, we plan to evaluate existing process mining techniques addressing privacy and confidentiality according to minimum loss of utility for process mining. Such an analysis would improve the selection of suitable techniques.


\bibliographystyle{ACM-Reference-Format}
\bibliography{literature}


\end{document}